\documentclass[12pt]{article}
\usepackage{graphicx}

\newcommand\pubnumber{SNSN-323-63}
\newcommand\pubdate{\today}

\def\napoli{Kamioka Observatory, Institute for Cosmic Ray Research,\\
University of Tokyo, Kamioka, Gifu 506-1205, Japan}
\def\Title#1{\begin{center} {\Large #1 } \end{center}}
\def\Author#1{\begin{center}{ \sc #1} \end{center}}
\def\Address#1{\begin{center}{ \it #1} \end{center}}

\newcommand\pubblock{\rightline{\begin{tabular}{l} \pubnumber\\
         \pubdate  \end{tabular}}}
\newenvironment{Abstract}{\begin{quotation}  }{\end{quotation}}
\newenvironment{Presented}{\begin{quotation} \begin{center} 
             PRESENTED AT\end{center}\bigskip 
      \begin{center}\begin{large}}{\end{large}\end{center} \end{quotation}}
\def\Acknowledgements{\bigskip  \bigskip \begin{center} \begin{large}
             \bf ACKNOWLEDGEMENTS \end{large}\end{center}}
\def\beq{\begin{equation}}
\def\eeq#1{\label{#1}\end{equation}}
\def\eeqn{\end{equation}}

\def\beqa{\begin{eqnarray}}
\def\eeqa#1{\label{#1}\end{eqnarray}}
\def\eeqan{\end{eqnarray}}

\let\bar=\overbar

\def\Dslash{\not{\hbox{\kern-4pt $D$}}}
\def\dslash{\not{\hbox{\kern-2pt $\del$}}}

\def\msb{{\bar{\ssstyle M \kern -1pt S}}}

\begin{document}
\begin{titlepage}
\pubblock

\vfill
\Title{Supernovae in SuperK-Gd and other experiments}
\vfill
%\Author{ Llu\'is Mart\'i Magro \support}
\Author{ Llu\'is Mart\'i Magro }
\Address{\napoli}
\vfill
\begin{Abstract}
Core-collapse supernovae are one of the most energetic events in the universe ($10^{46} J$). When a massive star (M $>$ 8 M$_{\odot}$) ignites its last fusion stage where silicon fusion makes iron, its end is then very close. Basically, the core of the star falls inwardly and the gravitational energy is then released in a supernova explosion. The basic picture of this explosion was confirmed by the few neutrinos detected from the SN1987a supernova at Kamiokande, IMB and Baksan detectors. However, there are many details that are still unknown. Since then, a large detector network has grown with better capabilities. Nowadays, in the case of a supernova explosion in our galaxy, the information that we would acquire would allow us to learn much more about these energetic events and constrain our models. Here, I present a brief summary of this network with special emphasis in SuperK-Gd (the upgraded Super-Kamiokande detector with efficient neutron tagging).
\end{Abstract}
\vfill
\begin{Presented}
NuPhys2016, Prospects in Neutrino Physics\\
Barbican Centre, London, UK, December 12-14, 2016
\end{Presented}
\vfill
\end{titlepage}
\def\thefootnote{\fnsymbol{footnote}}
\setcounter{footnote}{0}

\section{Introduction}

Core-collapse supernovae (ccSNe) are wonderful laboratories where we can learn a good deal of physics. The crucial importance of neutrinos and the basic mechanism on how massive stars undergo SN explosions was first described by~\cite{Colgate:1966ax, Arnett}. This mechanism was then confirmed by the two dozen neutrinos detected from the SN1987a SN explosion in the Large Magellanic Cloud on February 23, 1987: the source of their vast energy (from the gravitational collapse of the core of a massive star), the conversion of 99$\%$ of this energy into neutrinos and their mean temperature along with the explosion time duration (circa 12 seconds). After travelling 50 kpc, those neutrinos were detected at the Kamiokande II, IMB and Baksan detectors with 11, 8 and 5 neutrino events and fiducial masses of 2140 ton (water Cherenkov), 5000 ton (water Cherenkov) and 200 ton (liquid scintillator), respectively. These neutrinos were mainly detected from inverse beta decay (IBD) reactions: $\bar{\nu}_{e} + p \rightarrow n + e^+$. 

Despite the many efforts of several groups working in the theoretical details of ccSN explosions, the details are not fully understood and is still a very active research area. These groups try to simulate SN explosions but until now they have not been completely successful yet. They went from one dimensional computer simulations with perfect spherical symmetry to (still imperfect) three dimensional computer simulations with nonradial motions in the collapsing core of the star. In the early days, it was thought that the infall of the overlaying shells was stopped and then expelled by the bounce of this material on the forming neutron star. However, this bounce-shock mechanism has been replaced by a more successful yet not perfect delayed neutrino-heating mechanism with still many unknowns~\cite{Bethe:1984ux, Burrows:1995ww}.

\section{Current supernova detectors}

In parallel to these efforts in the theoretical front, a large network of detectors that would be able to detect SN neutrinos has been growing around the world. In the Kamioka mine (Japan) were Kamiokande was built are now operating two detectors: KamLAND and Super-Kamiokande (SuperK) while the Baksan detector (Northern Caucasus, Russia) is still running. 

Other neutrino detectors currently running are the Large Volume Detector (LVD) and Borexino in Gran Sasso (Italy), NOvA at Fermilab (USA), HALO at SNOLAB (Canada) and IceCube at the Amudsen-Scott South Pole station (Antartica). See Table~\ref{tab:detectorsList}. These detectors can be grossly divided by the detector material which in turn determines the main detection channel.

%%%%%%%%%%%%%%%%%%%%%%%%%%%%%%%%%%%%%%%%%%%%%%%%%%%%%%%%%%%%%%%%%%%%%%%%%
\begin{table}[t]
\begin{center}
\begin{tabular}{l|cccc}  
Detector          &  Mass      &  Detector material    & Operation begin   & Country  \\ \hline
 KamLAND          &  1 kt        &  liquid scintillator  &  2002           & Japan  \\
 Super-Kamiokande &  32 kt       &  ultra-pure water     &  1996           & Japan  \\ 
 Baksan           &  0.3 kt      &  liquid scintillator  &  1980           & Russia \\ 
 LVD              &  1 kt        &  liquid scintillator  &  1992           & Italy  \\ 
 Borexino         &  0.3 kt      &  liquid scintillator  &  2007           & Italy  \\ 
 NOvA             &  14 kt       &  liquid scintillator  &  2014           & USA    \\ 
 HALO             &  76 t        &  lead                 &  2010           & Canada \\ 
 IceCube          &  1 Gt        &  antartic ice         &  2005           & (Antartica) \\ \hline
\end{tabular}
\caption{List of currently operating SN neutrino detectors.}
\label{tab:detectorsList}
\end{center}
\end{table}
%%%%%%%%%%%%%%%%%%%%%%%%%%%%%%%%%%%%%%%%%%%%%%%%%%%%%%%%%%%%%%%%%%%%%%%%%%%

\subsection{Liquid scintillator detectors}

Liquid scintillator (C$_n$H$_{2n}$) detectors (LSD) have a low energy threshold (below 1 MeV) and have good energy resolution. These detectors can also use the delayed coincidence neutron tagging technique, which is very useful in IBD. An important drawback of LSDs is that they basically offer no SN directionality information since the produced light is almost isotropic. The main interaction channels and the rough number of neutrino events that are expected in these detectors in case of a SN at 10 kpc is shown in Table~\ref{tab:lcd_channels}.

%%%%%%%%%%%%%%%%%%%%%%%%%%%%%%%%%%%%%%%%%%%%%%%%%%%%%%%%%%%%%%%%%%%%%%%%%
\begin{table}[t]
\begin{center}
\begin{tabular}{l|ccc}  
Interaction channel                                             &  Events/kton &  Comments      \\ \hline
 $\bar{\nu}_{e} + p \rightarrow n + e^+$                        &  300         &  spectrum distortion from earth matter effect\\
 $\nu_{e} + C^{12} \rightarrow e + N^{12} (B^{12})$             &  30          &  tagged by $N^{12} (B^{12})$ beta decay   \\ 
 $\nu + C^{12} \rightarrow \nu + C^{12} + \gamma (15.1 MeV) $   &  60          &  no pointing capability in LSDs \\ 
 $\nu + e^- \rightarrow \nu + e^- $                             &  20          &  all flavours (insensitive to oscillations) \\ 
 $\nu + p \rightarrow \nu + p $                                 &  300         &  all flavours (higher energy component) \\ \hline
\end{tabular}
\caption{Main interaction channels in liquid scintillator detectors. Events/kton for a SN at the galactic center (10 kpc) are shown.}
\label{tab:lcd_channels}
\end{center}
\end{table}
%%%%%%%%%%%%%%%%%%%%%%%%%%%%%%%%%%%%%%%%%%%%%%%%%%%%%%%%%%%%%%%%%%%%%%%%%%%

\begin{figure}[htb]
\centering
\includegraphics[height=2.in]{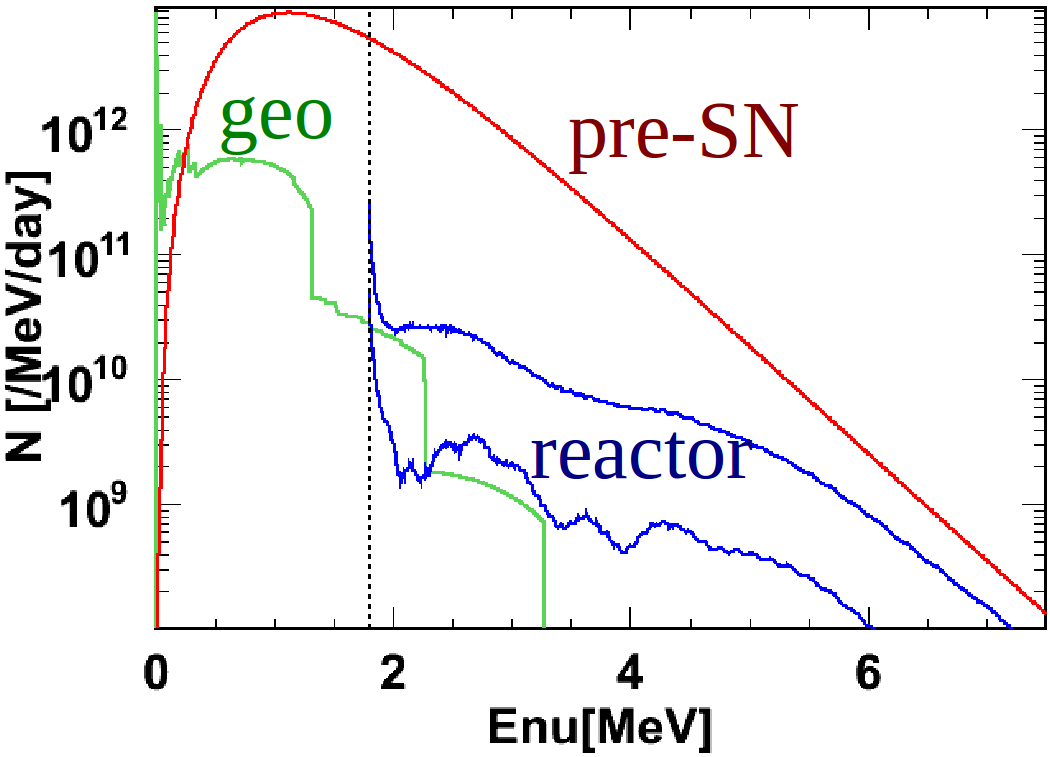}
\caption{Expected neutrino spectrum from a pre-SN star at 200 pc and compared to geo- and reactor neutrino rates in KamLAND.}
\label{fig:KamLAND_pre-SN}
\end{figure}

Massive stars in their late stages of their evolution are referred as neutrino-cooled stars. During the C, Ne, O and Si fusion phases they emit vast amounts of neutrinos at an increasing rate. In fact, in the last phase of Si fusion, the production of neutrino-antineutrino pairs is such that KamLAND could detect a pre-SN star if this would be close enough~\cite{Asakura:2015bga}. In Figure~\ref{fig:KamLAND_pre-SN} it is shown the expected spectrum of these neutrinos for a star at 200 pc and compared to geo- and reactor neutrino rates. Given the enormous advantages the detection of a pre-SN represents, KamLAND has prepared a pre-SN alarm system that would be sensitive for stars up to 700 pc.

%%%%%%%%%%%%%%%%%%%%%%%%%%%%%%%%%%%%%%%%%%%%%%%%%%%%%%%%%%%%%%%%%%%%%%%%%
%%%%%%%%%%%%%%%%%%%%%%%%%%%%%%%%%%%%%%%%%%%%%%%%%%%%%%%%%%%%%%%%%%%%%%%%%%%

\subsection{Lead detectors}

The HALO detector is a quite unique SN neutrino detector~\cite{Zuber:2015ita}. HALO prioritizes a detector long lifetime and low construction and detector maintenance costs. It consists of about 80 tons of lead as target material. This lead was taken from the decommissioning of the Deep River Cosmic Ray Station and instrumented with He$^{3}$ neutron counters from the SNO 3rd phase. Neutrons are moderated in polypropylene and the detector runs with SNO electronics. Lead is a neutron rich material and thus, it offers a good sensitivity to $\nu_e$'s through charged current (CC) processes (p $\rightarrow$ n transitions are here suppressed). This is in clear contrast to water Cherenkov and LSDs, which are primarily sensitive to $\bar{\nu}_e$'s through IBD. HALO is also sensitive to all six $\nu$ and $\bar{\nu}$ species through neutral current interactions (NC). For a SN at the centre of the galaxy (10 kpc) the expected number of neutrons produced is about 88 neutrons, see Table~\ref{tab:halo}. With an efficiency of about 43$\%$, HALO would detect about 40 neutrons. There is no charged current/neutral current (CC/NC) event distinction but the ratio of 1n/2n events may yield spectral information of the SN neutrino flux. There are plans for HALO-2, a kton-scale lead detector.% [refs].

%%%%%%%%%%%%%%%%%%%%%%%%%%%%%%%%%%%%%%%%%%%%%%%%%%%%%%%%%%%%%%%%%%%%%%%%%
\begin{table}[t]
\begin{center}
\begin{tabular}{l|ccc}
Interaction channel                                             &  Events &  Interaction            \\ \hline
 $\nu_e + Pb^{208} \rightarrow Bi^{207} + n + e^- $             &  30     &  charged current  \\ 
 $\nu_e + Pb^{208} \rightarrow Bi^{206} + 2n + e^- $            &  19     &  charged current  \\ 
 $\nu_x + Pb^{208} \rightarrow Pb^{207} + n + e^- $             &  8      &  neutral current \\
 $\nu_x + Pb^{208} \rightarrow Pb^{206} + 2n + e^- $            &  6      &  neutral current \\ \hline
\end{tabular}
\caption{Number of expected neutrinos events for a SN at the centre of the galaxy (10 kpc) for the HALO detector.}
\label{tab:halo}
\end{center}
\end{table}
%%%%%%%%%%%%%%%%%%%%%%%%%%%%%%%%%%%%%%%%%%%%%%%%%%%%%%%%%%%%%%%%%%%%%%%%%%%

\subsection{Water Cherenkov detectors}

\begin{figure}[htb]
\centering
\includegraphics[height=2.in]{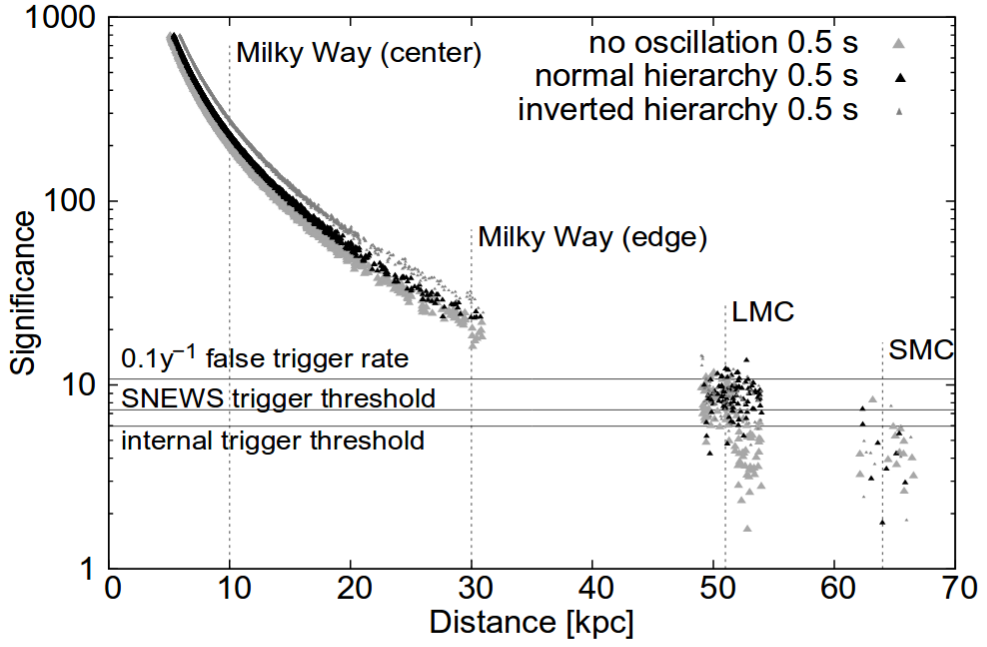}
\caption{IceCube SN detection significance as a function of distance in case of no oscillation, normal and inverted hierarchy and for a 0.5 seconds binning (Lawrence-Livermore model). The Milky Way as well as the Large and the Small Magellanic Cloud (LMC and SMC) are included.}
\label{fig:IceCube}
\end{figure}

Water Cherenkov detectors have many good features for SN neutrino detection: water is an abundant, convenient and cheap detector material. Thus, despite the light yield is lower than in LSDs, very large instrumented volumes are relatively easy to deploy. IceCube is a Giga-ton detector with kilometre long PMT strings deployed in the ice of the South Pole. Designed for multi-GeV neutrinos, it can detect a low energy $\bar{\nu}_e$ burst as an increase in the single PMT count rate. In case of a SN neutrino burst the ice would be uniformly illuminated. Therefore, by detecting the correlated increase of single PMT count rates on top of the PMT noise, IceCube can detect subtle temporal features with high statistics~\cite{Abbasi:2011ss}, see Figure~\ref{fig:IceCube}. However, IceCube cannot deliver any SN pointing or neutrino energy spectrum information.

%%%%%%%%%%%%%%%%%%%%%%%%%%%%%%%%%%%%%%%%%%%%%%%%%%%%%%%%%%%%%%%%%%%%%%%%%
%%%%%%%%%%%%%%%%%%%%%%%%%%%%%%%%%%%%%%%%%%%%%%%%%%%%%%%%%%%%%%%%%%%%%%%%%%%

\begin{figure}[htb]
\centering
\includegraphics[height=3.5in]{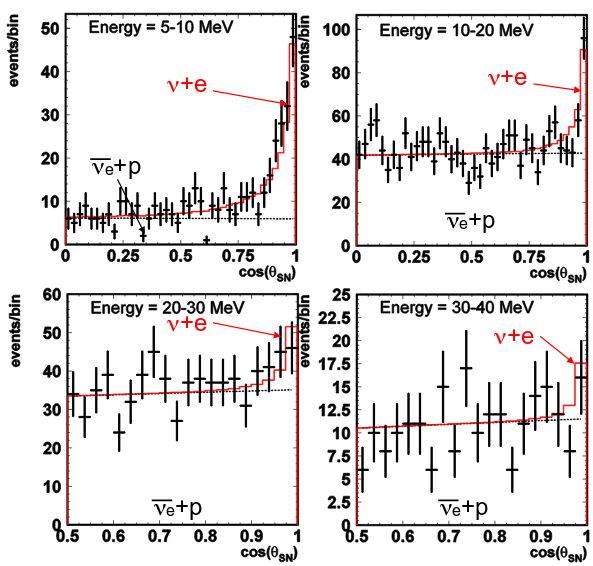}
\caption{SuperK can determine the SN direction with an accuracy of 4-5$^{\circ}$ for a SN at 10 kpc.}
\label{fig:SuperK-directionality}
\end{figure}

SuperK is a 50 kton ultra-pure water detector in Kamioka (Japan). The detector is divided in an outer detector that acts as a veto for cosmic rays and an inner detector of 32 kton viewed by $\sim$11.100 20-inch PMTs with a 4.5 MeV kinetic energy threshold. As for LSDs, the most important interaction is IBD, see Table~\ref{tab:superK}. However, electron elastic scattering events carry information about the position of the SN in the sky, which is a unique capability among all SN detectors. Being $\hat{d}_i$ the direction of the $i$-th event and $\hat{d}_{SN}$ the SN direction, then we can define cos $\theta_{SN} = \hat{d}_i \cdot \hat{d}_{SN}$. In Figure~\ref{fig:SuperK-directionality}, cos $\theta_{SN}$ is shown for MC events in the following energy ranges: 5-10 MeV, 10-20 MeV, 20-30 MeV and 30-40 MeV. With the elastic scattering events, SuperK can determine the direction of a SN at 10 kpc with an accuracy of 4-5$^{\circ}$. From this figure it is clear that if SuperK could remove IBD events the pointing accuracy would be greatly improved.

%%%%%%%%%%%%%%%%%%%%%%%%%%%%%%%%%%%%%%%%%%%%%%%%%%%%%%%%%%%%%%%%%%%%%%%%%
\begin{table}[t]
\begin{center}
\begin{tabular}{l|ccc}
Interaction channel                                             &  Events &  Interaction                       \\ \hline
 $\bar{\nu}_{e} + p \rightarrow n + e^+$                        &  7300   &  IBD                               \\
 $\nu_x + O^{16} \rightarrow \nu_x + O^{16} + \gamma $          &  360    &  $O^{16}$ neutral current $\gamma$ \\ 
 $\nu_x + e^- \rightarrow \nu_x + e^- $                         &  320    &  elastic scattering                \\
 $\nu_e + O^{16} \rightarrow e + F^{16} (N^{16})$               &  100    &  $O^{16}$ charged current          \\ \hline
\end{tabular}
\caption{Number of expected neutrino events for a SN at the centre of the galaxy (10 kpc) with a 4.5 MeV kinetic energy threshold (Livermore simulation).}
\label{tab:superK}
\end{center}
\end{table}
%%%%%%%%%%%%%%%%%%%%%%%%%%%%%%%%%%%%%%%%%%%%%%%%%%%%%%%%%%%%%%%%%%%%%%%%%%%

%%%%%%%%%%%%%%%%%%%%%%%%%%%%%%%%%%%%%%%%%%%%%%%%%%%%%%%%%%%%%%%%%%%%%%%%%
%%%%%%%%%%%%%%%%%%%%%%%%%%%%%%%%%%%%%%%%%%%%%%%%%%%%%%%%%%%%%%%%%%%%%%%%%%%

To achieve this goal, neutron tagging is needed at SuperK. Currently, IBD neutrons are captured on protons after $\sim$200 $\mu$s and a 2.2 MeV gamma is then produced. The problem is that this is not efficiently detected. GADZOOKS!~\cite{Beacom:2003nk} was proposed to achieve neutron tagging with high efficiency. The basic idea is to dope with gadolinium (Gd) the otherwise SuperK ultra-pure water. This is done by dissolving Gd sulfate, Gd$_2$(SO$_4$)$_3$. Gd has a thermal neutron capture cross-section of 49000 barn (to be compared to that of the proton of 0.33 barn). With only 0.2$\%$ of Gd sulfate (0.1 $\%$ of Gd) 90$\%$ of the captures will be on Gd. After the neutron capture (capture time of $\sim$30 $\mu$s), a gamma cascade of 8 MeV and shared among 3-4 gammas is produced which can be detected with high efficiency.

The initial motivation for GADZOOKS! was the diffuse SN neutrino background (DSNB), i.e. the neutrinos from all the past ccSN in the history of the visible universe. Although SuperK has the best world limit on DSNB~\cite{Bays:2011si} large irreducible backgrounds dominate this search. However, by adding neutron tagging capabilities to SuperK these irreducible backgrounds could be greatly reduced and within 10 years of observation DSNB would be within our reach (with a significance that would depend in the typical SN emission spectrum).

\begin{figure}[htb]
\centering
\includegraphics[height=3.5in]{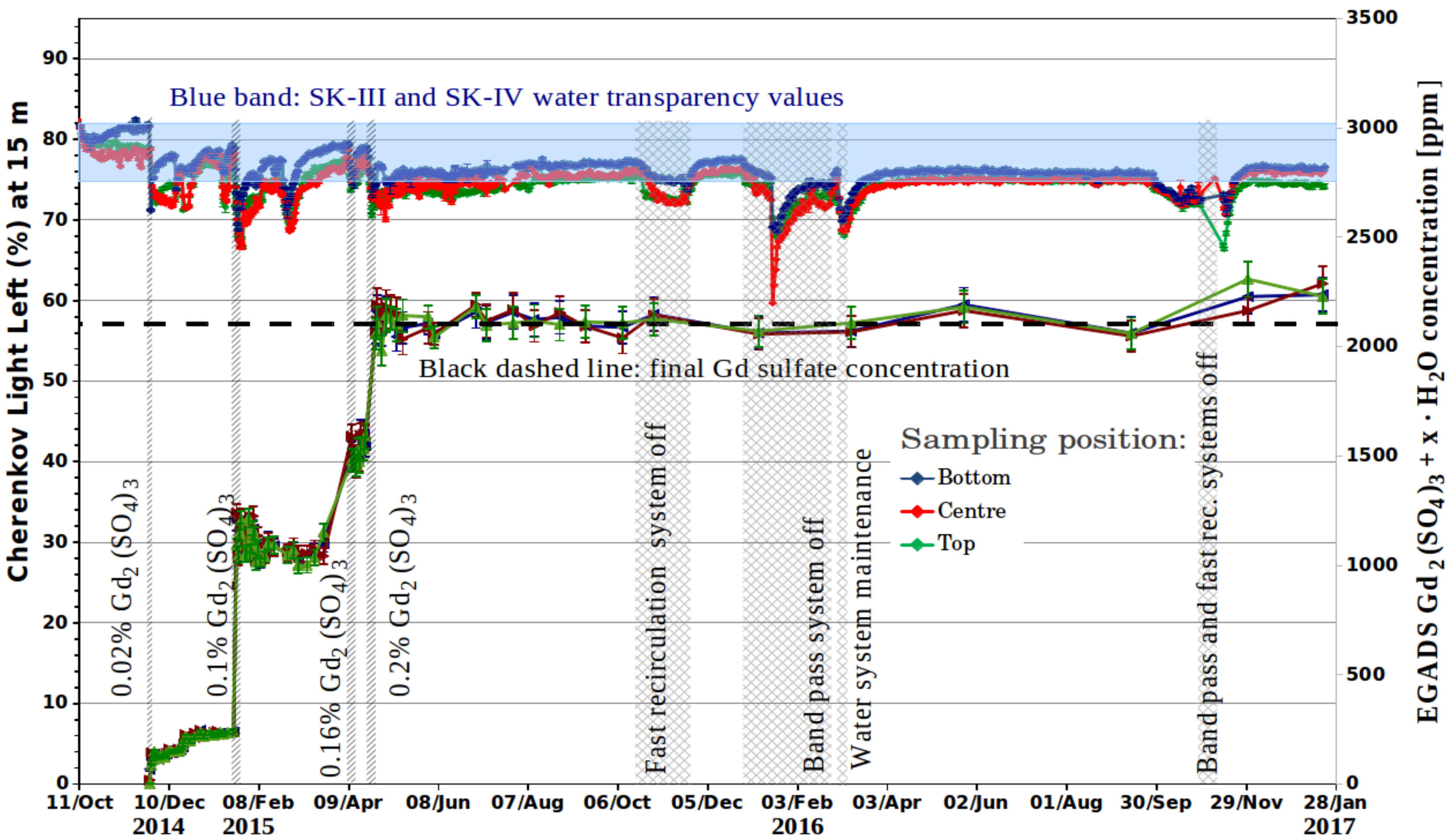}
\caption{Cherenkov light left after 15 m (LL15) is shown for three different sampling positions in the EGADS detector. The blue band indicates typical values for SuperK III and IV. In the same figure, Gd sulfate concentration for the same sampling points is shown. The black dashed line indicates the final Gd sulfate concentration while the vertical bands indicate the Gd sulfate loadings and other experimental events.}
\label{fig:EGADS_results}
\end{figure}

In 2009 the R$\&$D project EGADS (Evaluating Gadolinium's Action on Detector Systems) was funded in Japan and a new hall in the Kamioka mine near SuperK was excavated. The purpose of the project was to demonstrate the feasibility of GADZOOKS!. EGADS features a 200-ton detector with 240 photo-detectors with its own water purification system (specially designed to remove all impurities in water but to keep Gd). In EGADS, all materials were chosen of the same type as in SuperK in order to mimic the conditions there. On June 27, 2015, the Super-Kamiokande collaboration approved the SuperK-Gd project. In Figure~\ref{fig:EGADS_results}, the Cherenkov light left after 15 m (LL15) is shown for three different sampling positions in the EGADS detector. The blue band indicates typical values for SuperK III and IV. In the same figure, Gd sulfate concentration for the same sampling points is shown. The black dashed line indicates the final Gd sulfate concentration while the vertical bands indicate the Gd sulfate loadings and other experimental events. After the last Gd sulfate loading, the measured LL15 values in EGADS are inside the blue band when running in stable conditions while the Gd sulfate concentration remains stable.

%%%%%%%%%%%%%%%%%%%%%%%%%%%%%%%%%%%%%%%%%%%%%%%%%%%%%%%%%%%%%%%%%%%%%%%%%
%%%%%%%%%%%%%%%%%%%%%%%%%%%%%%%%%%%%%%%%%%%%%%%%%%%%%%%%%%%%%%%%%%%%%%%%%%%

With efficient neutron tagging, the IBD events could be removed from the cos~$\theta_{SN}$ distributions, see Figure~\ref{fig:EGADS_results}. As a consequence, the direction of a galactic SN would be improved. For a SN at 10 kpc it could be determined with an accuracy of about 3$^{\circ}$.

%%%%%%%%%%%%%%%%%%%%%%%%%%%%%%%%%%%%%%%%%%%%%%%%%%%%%%%%%%%%%%%%%%%%%%%%%
\begin{figure}[htb]
\centering
\includegraphics[height=3.5in]{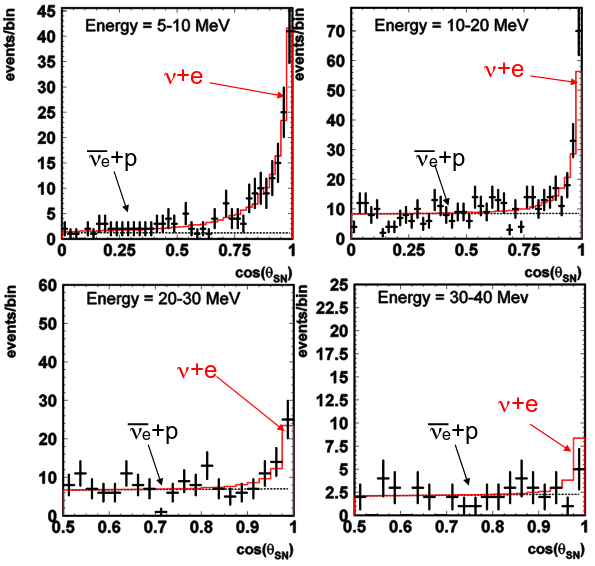}
\caption{With an efficient neutron tagging, SuperK SN direction accuracy would be improved from about 4-5$^{\circ}$ to about 3$^{\circ}$ for a SN at 10 kpc.}
\label{fig:SuperK-directionality_Gd}
\end{figure}
%%%%%%%%%%%%%%%%%%%%%%%%%%%%%%%%%%%%%%%%%%%%%%%%%%%%%%%%%%%%%%%%%%%%%%%%%%%

\section{Future supernova detectors}

A new generation of future detectors is underway. In the next years, we will see new Cherenkov detectors like Hyper-Kamiokande, PINGU and IceCube-Gen2. Hyper-Kamiokande will consist of two tanks of 0.5 Mton total mass and 40 $\%$ photo-cathode coverage (40000 PMTs). New 50-cm PMTs are being developed with better detection efficiency and timing resolution. Because they will be installed in deeper tanks, these PMTs also designed to have a better pressure tolerance. PINGU and IceCube-Gen2 will represent a low-energy infill extension for IceCube and a substantial expansion to about 10 km$^3$ of the current IceCube detector, respectively. This will greatly improve IceCube galactic SN sensitivity among other physics goals.

DUNE, a 40 kton liquid Ar detector, is planed to be deployed in the Homestake mine (South Dakota). Designed to be sensitive to neutrinos in the few tens of MeV range and in particular to the $\nu_e$ component of a SN neutrino burst via CC.
%  ~\cite{Mesmer}. \index{Mesmer}     

A 20 kton LSD has also been proposed. JUNO, designed for a rich physics program is designed to achieve an excellent energy resolution and a large fiducial volume. For a SN at 10 kpc, JUNO will see $\sim$5000 IBD events and $\sim$2000 all-flavour neutrino-proton elastic scattering events. Because of its neutron tagging capabilities it also aims to study the DSNB.

\Acknowledgements
I would like to thank the organizing committee for the invitation to the NuPhys2016 conference and the fruitful discussions.

\end{document}